\documentclass[11pt,twoside]{article}

\usepackage{asp2014}

\aspSuppressVolSlug
\resetcounters

\begin{document}

\title{Machine Learning Workflow for Morphological Classification of Galaxies}

\author{Bernd Doser, Kai L. Polsterer, Andreas Fehlner and Sebastian Trujillo-Gomez}
\affil{Heidelberg Institute for Theoretical Studies, Heidelberg, Germany; \email{bernd.doser@h-its.org}}

\paperauthor{Bernd Doser}{bernd.doser@h-its.org}{0000-0002-3443-5913}{HITS gGmbH}{Astroinformatics}{Heidelberg}{BW}{69118}{Germany}
\paperauthor{Kai Lars Polsterer}{kai.polsterer@h-its.org}{https://orcid.org/0000-0002-3435-1912}{HITS gGmbH}{Astroinformatics}{Heidelberg}{BW}{69118}{Germany}
\paperauthor{Andreas Fehlner}{andreas.fehlner@h-its.org}{0000-0003-0448-7138}{HITS gGmbH}{Astroinformatics}{Heidelberg}{BW}{69118}{Germany}
\paperauthor{Sebastian Trujillo-Gomez}{sebastian.trujillogomez@h-its.org}{0000-0003-2482-0049}{HITS gGmbH}{Astroinformatics}{Heidelberg}{BW}{69118}{Germany}



\begin{abstract}
As part of the EU-funded Center of Excellence SPACE (Scalable Parallel Astrophysical Codes for Exascale), seven commonly used astrophysics simulation codes are being optimized to exploit exascale computing platforms.
Exascale cosmological simulations will produce large amounts of data (i.e. several petabytes) that will soon be waiting to be analyzed, with enormous potential for scientific breakthroughs. Our tool Spherinator \citep{Polsterer_2024} enables the reduction of these complex data sets to a low-dimensional space using Generative Deep Learning to understand the morphological structure of simulated galaxies.
A spherical latent space allows the HiPSter module to provide explorative visualization using Hierarchical Progressive Surveys (HiPS) \citep{Fernique_2015} in the Aladin software.

Here we present a machine-learning workflow covering all stages, from data collection to preprocessing, training, prediction, and final deployment.
This workflow ensures full reproducibility by keeping track of the code, data, and environment. Additionally, the workflow allows for scalability in managing a large amount of data and complex pipelines.
We use only open source software and standards that align with the FAIR (Findability, Accessibility, Interoperability and Reproducibility) principles \citep{Wilkinson_2016}.
In this way, we are able to distribute the workflow reliably and enable collaboration by sharing code, data, and results efficiently.
\end{abstract}

\section{Introduction}

Cosmological simulations typically output an enormous amount of data, with the largest ones exceeding $\sim 1$ PB of data. The outputs are not only large, but high-dimensional: They include the detailed physical properties of millions of structures composed of billions of particles, each described by dozens of features (including position, velocity, density, temperature, metallicity and individual element abundances, among many others \citep{Nelson_2015}).
Enhanced techniques and methods are required to process data of this size and complexity efficiently.
Machine learning offers a powerful way to reduce the dimensionality of simulation data and ensure at the same time that similar objects maintain a comparable compressed representation.

We have developed two methods for dimensionality reduction.
PINK \citep{PINK_2016} is a previous development that utilizes a self-organizing Kohonen map to classify radio galaxy maps.
In the framework of the EU-funded Center of Excellence SPACE, we have developed Spherinator \citep{Polsterer_2024}, which employs a variational autoencoder with a convolutional neural network to create an explorable 2D representation of simulated galaxy images.
Here we present a comprehensive workflow to facilitate integrating current and emerging dimensionality reduction techniques.
This workflow takes the FAIR principles into account, ensuring that these foundational guidelines are appropriately addressed.
We present two distinct workflow orchestration frameworks: one specifically designed for high-performance computing (HPC) clusters and one optimized for cloud-native environments.

\section{The Workflow}

\articlefigure{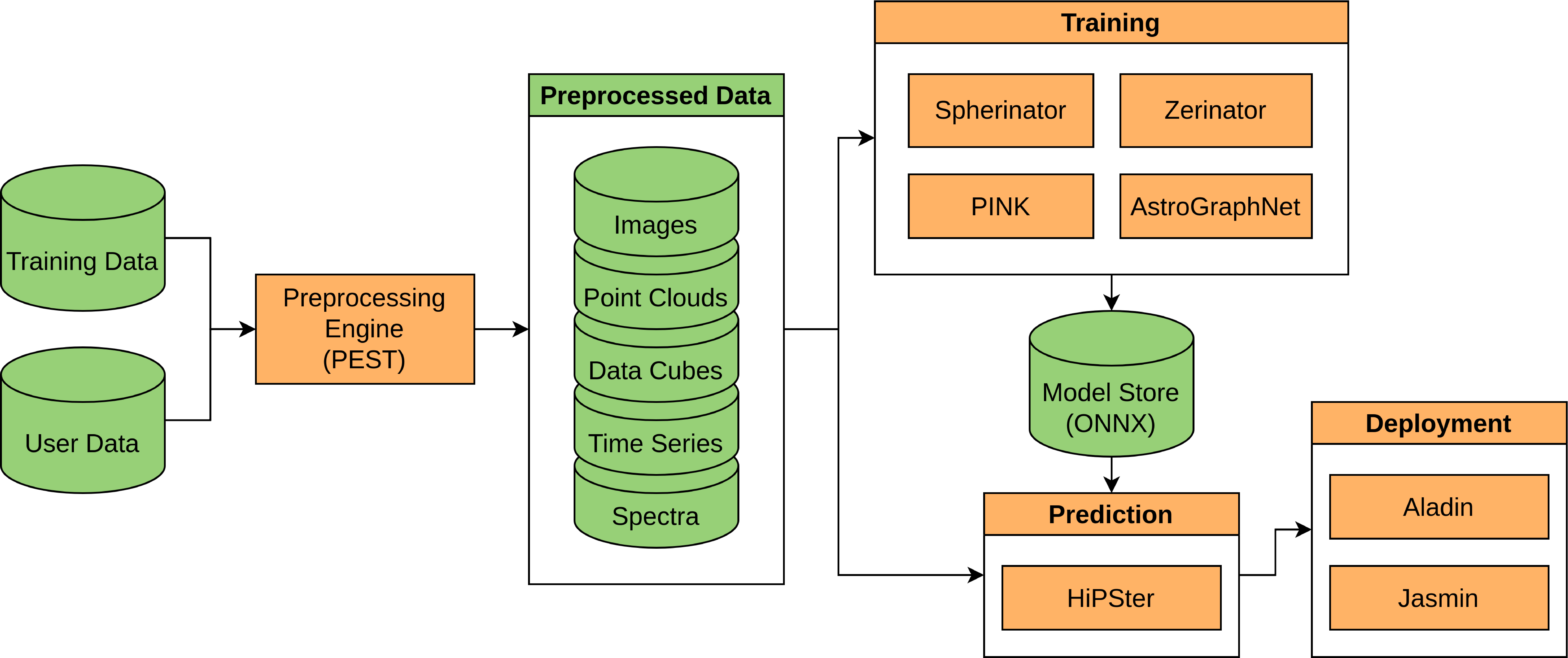}{P201_fig1}{Schematic workflow for training and inference using a compact representation of galaxy morphology.}

The workflow (see Figure~\ref{P201_fig1}) is designed with an adaptable structure, featuring general modules for data preprocessing, model training, prediction (model inference), and deployment, making it suitable for various tasks.
As a first step, the \textit{Preprocessing Engine for Spherinator Training} (PEST) processes many types of data, including astrophysical simulations and observations, into universal representations regardless of their origin.
PEST converts raw data into a unique dataset optimized for machine learning model training and inference.
The module uses Apache Parquet as a data format optimized for reading and writing data frames due to its columnar storage structure.
The model training can use a deep learning variational autoencoder (Spherinator), self-organizing Kohonen maps (PINK), or any other algorithm that may become available in the future, e.g. involving Zernike polynomials or graph neural networks (GNNs).
The trained models are exported into the ONNX (Open Neural Network Exchange) format, an open standard for compatibility with various frameworks and programming languages.
The prediction module performs the inference using the trained model. For example, the HiPSter module projects the spherical latent space of the Spherinator module onto HiPS tiles.
The final deployment module performs the user interaction using Aladin Lite to visualize the hierarchical spherical latent space, and Jasmin to display 3D structures interactively.

\section{Workflow Orchestration}

\articlefigure[width=.9\textwidth]{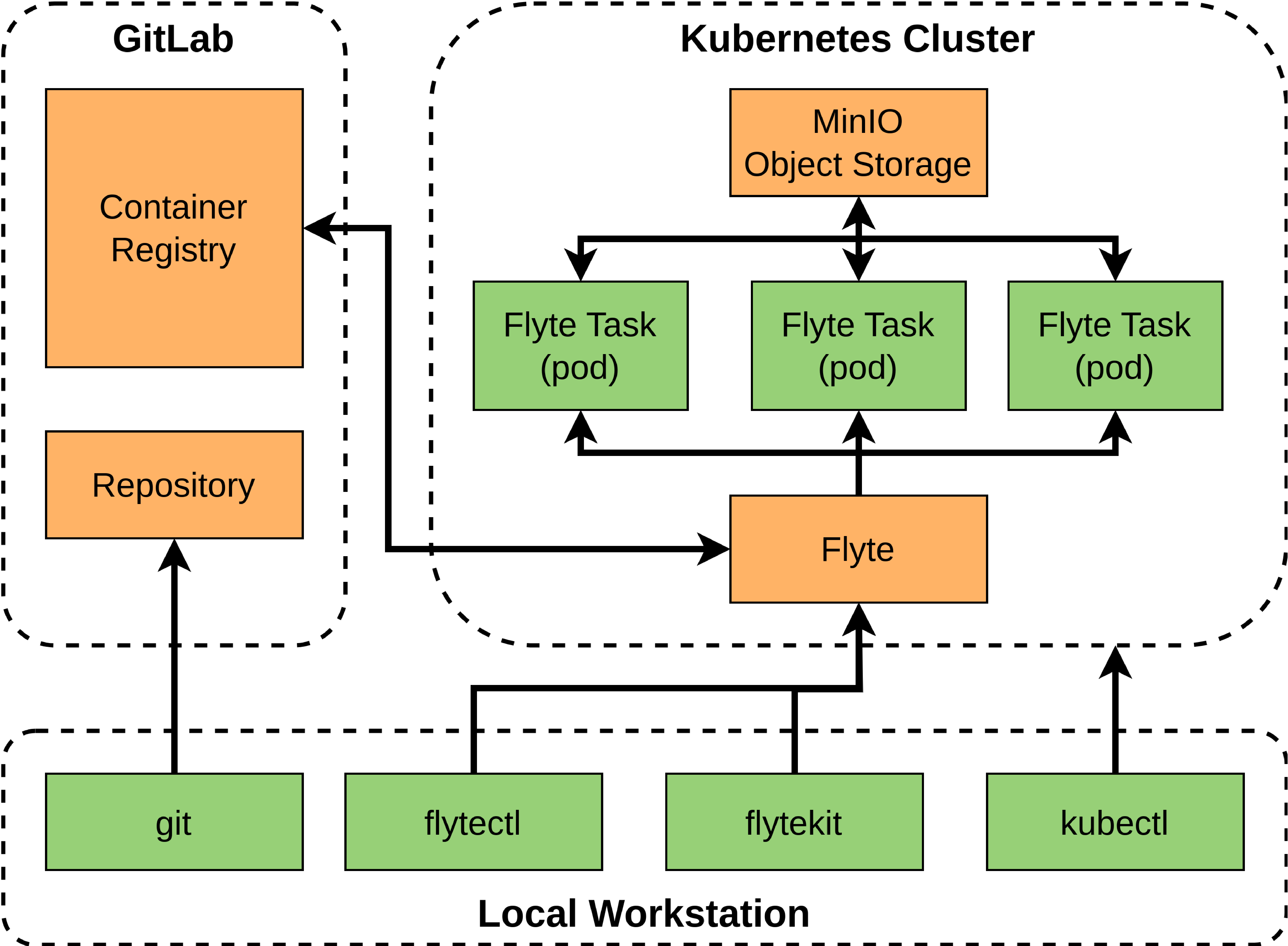}{P201_fig2}{Setup of the workflow orchestration cluster using the Flyte cloud-native platform.}

Workflow orchestration aims to define and execute workflows and data pipelines on a cloud-native or high-performance computing (HPC) cluster.
However, since the operational methods of HPC clusters (typically managed by Slurm) and cloud-native clusters (typically managed by Kubernetes) differ fundamentally, we have developed two separate orchestration frameworks adapted for each case.

StreamFlow\footnote{\url{https://github.com/alpha-unito/streamflow}} is available as a Python package and has a simple installation process that makes workflow distribution easy. The relationships between tasks are defined using the Common Workflow Language (CWL), and data is mainly exchanged through a distributed file system.
StreamFlow provides an ideal solution for HPC clusters, where the compute nodes can only be used via Slurm.

Flyte\footnote{\url{https://github.com/flyteorg/flyte}} is a highly scalable, cloud-native workflow orchestration platform that\linebreak leverages containers, Kubernetes, and object storage to facilitate data pipelines.
Figure~\ref{P201_fig2} provides a schematic overview of the Flyte architecture. The Flyte platform, which encompasses the administration interface, scheduler, and dashboard, is integrated within a Kubernetes cluster. Workflows are registered through a command line interface provided by Flyte. Each task within the workflow is executed in a dedicated Kubernetes pod and container, and data is transferred between tasks via object storage.

The requirements of Flyte to the environment are more complex than those of StreamFlow, but this also allows for a more flexible approach to defining workflows. Flyte's data pipeline, which utilizes Python's static type system and object storage, offers significantly more stability than the StreamFlow solution.

\section{Conclusions}

Workflows and their orchestration are of critical importance to modern data-driven astrophysics and cosmology in several respects as they enable:
\begin{itemize}
  \item Reproduciility of  computations by tracking the code, data, models, and environments.
  \item Efficient compute resource management by running tasks in parallel and optimizing resource usage.
  \item Pipelines to be scaled up to handle large amounts of data and complex pipelines
  \item Tracking workflow progress and monitoring performance and results.
  \item Facilitating sharing code, data, and results to improve collaboration.
\end{itemize}

In this work we introduced a workflow for dimensionality reduction techniques.
A key component of the workflow is a novel universal dataset format utilizing Apache Parquet for storing multiple data types in a data loader-friendly columnar storage structure.
In addition we presented two workflow orchestration frameworks:
While StreamFlow is, in principle, applicable to both cloud-native and HPC clusters, we have used it to build an orchestration for HPC and Flyte for cloud-native environments.
Our developments are broadly applicable beyond the explorative visualization of galaxy morphologies.

\acknowledgements 

We gratefully acknowledge the generous and invaluable support of the Klaus Tschira Foundation.
This work has received funding from the European High Performance Computing Joint Undertaking (JU) and Belgium, Czech Republic, France, Germany, Greece, Italy, Norway, and Spain under grant agreement No101093441.
Views and opinions expressed are however those of the author(s) only and do not necessarily reflect those of the European Union or the European High Performance Computing Joint Undertaking (JU) and Belgium, Czech Republic, France, Germany, Greece, Italy, Norway, and Spain.
Neither the European Union nor the granting authority can be held responsible for them.
All code is available at \url{https://github.com/HITS-AIN/Spherinator}.
We thank Martin Wendt for setting up the cluster using Kubernetes, Flyte, and MinIO.

\bibliography{P201}  

\begin{thebibliography}{}
\expandafter\ifx\csname natexlab\endcsname\relax\def\natexlab#1{#1}\fi
\expandafter\ifx\csname url\endcsname\relax
  \def\url#1{\texttt{#1}}\fi
\expandafter\ifx\csname urlprefix\endcsname\relax\def\urlprefix{URL }\fi
\providecommand{\eprint}[2][]{\url{#2}}

\bibitem[{{Fernique} et~al.(2015){Fernique}, {Allen}, {Boch}, {Oberto},
  {Pineau}, {Durand}, {Bot}, {Cambr{\'e}sy}, {Derriere}, {Genova}, \&
  {Bonnarel}}]{Fernique_2015}
{Fernique}, P., {Allen}, M.~G., {Boch}, T., {Oberto}, A., {Pineau}, F.~X.,
  {Durand}, D., {Bot}, C., {Cambr{\'e}sy}, L., {Derriere}, S., {Genova}, F., \&
  {Bonnarel}, F. 2015, \aap, 578, A114. \eprint{1505.02291}

\bibitem[{Nelson et~al.(2015)Nelson, Pillepich, Genel, Vogelsberger, Springel,
  Torrey, Rodriguez-Gomez, Sijacki, Snyder, Griffen, Marinacci, Blecha, Sales,
  Xu, \& Hernquist}]{Nelson_2015}
Nelson, D., Pillepich, A., Genel, S., Vogelsberger, M., Springel, V., Torrey,
  P., Rodriguez-Gomez, V., Sijacki, D., Snyder, G., Griffen, B., Marinacci, F.,
  Blecha, L., Sales, L., Xu, D., \& Hernquist, L. 2015, Astronomy and
  Computing, 13, 12–37.
  \urlprefix\url{http://dx.doi.org/10.1016/j.ascom.2015.09.003}

\bibitem[{Polsterer et~al.(2024)Polsterer, Doser, Fehlner, \&
  Trujillo-Gomez}]{Polsterer_2024}
Polsterer, K.~L., Doser, B., Fehlner, A., \& Trujillo-Gomez, S. 2024,
  Spherinator and hipster: Representation learning for unbiased knowledge
  discovery from simulations. \eprint{2406.03810},
  \urlprefix\url{https://arxiv.org/abs/2406.03810}

\bibitem[{Polsterer et~al.(2016)Polsterer, Gieseke, Igel, Doser, \&
  Gianniotis}]{PINK_2016}
Polsterer, K.~L., Gieseke, F., Igel, C., Doser, B., \& Gianniotis, N. 2016, in
  ESANN 2016

\bibitem[{Wilkinson et~al.(2016)Wilkinson, Dumontier, Aalbersberg, Appleton,
  Axton, Baak, Blomberg, Boiten, da~Silva~Santos, Bourne
  et~al.}]{Wilkinson_2016}
Wilkinson, M.~D., Dumontier, M., Aalbersberg, I.~J., Appleton, G., Axton, M.,
  Baak, A., Blomberg, N., Boiten, J.-W., da~Silva~Santos, L.~B., Bourne, P.~E.,
  et~al. 2016, Scientific data, 3

\end{thebibliography}


\end{document}